# The mechanical response of cellular materials with spinodal topologies

Meng-Ting Hsieh, Bianca Endo, Yunfei Zhang, Jens Bauer, Lorenzo Valdevit[*]

*Mechanical and Aerospace Engineering Department, University of California, Irvine, CA 92697*

**Abstract**

The mechanical response of cellular materials with spinodal topologies is numerically and experimentally investigated. Spinodal microstructures are generated by the numerical solution of the Cahn-Hilliard equation. Two different topologies are investigated: 'solid models,' where one of the two phases is modeled as a solid material and the remaining volume is void space; and 'shell models,' where the interface between the two phases is assumed to be a solid shell, with the rest of the volume modeled as void space. In both cases, a wide range of relative densities and spinodal characteristic feature sizes are investigated. The topology and morphology of all the numerically generated models are carefully characterized to extract key geometrical features and ensure that the distribution of curvatures and the aging law are consistent with the physics of spinodal decomposition. Finite element meshes are generated for each model, and the uniaxial compressive stiffness and strength are extracted. We show that while solid spinodal models in the density range of 30-70% are relatively inefficient (i.e., their strength and stiffness exhibit a high-power scaling with relative density), shell spinodal models in the density range of 0.01-1% are exceptionally stiff and strong. Spinodal shell materials are also shown to be remarkably imperfection insensitive. These findings are verified experimentally by in-situ uniaxial compression of polymeric samples printed at the microscale by Direct Laser Writing (DLW). At low relative densities, the strength and stiffness of shell spinodal models outperform those of most lattice materials and approach theoretical bounds for isotropic cellular materials. Most importantly, these materials can be produced by self-assembly techniques over a range of length scales, providing unique scalability.

---

[*] Corresponding Author. E-mail: Valdevit@uci.edu

## 1. Introduction

Cellular materials are porous materials consisting of one or more solid phases and void space. For all cellular materials, Young's modulus, $E$ and yield (or buckling) strength, $\sigma_y$, are a strong function of the relative density, $\bar{\rho}$ (i.e., the volume fraction of the solid phase), generally following a power-law behavior: $E \sim E_s \bar{\rho}^n$ and $\sigma_y \sim \sigma_{ys} \bar{\rho}^m$, with $E_s$ and $\sigma_{ys}$ the Young's modulus and yield (or buckling) strength of the base material, respectively. The values of the exponents $n$ and $m$ have a dramatic effect on the mechanical efficiency of the material and are strongly affected by the topology of the unit cell architecture (Fleck et al., 2010).

The mechanical properties of open-cell, strut-based cellular materials like foams and lattices have been studied extensively (Ashby et al., 2000; Deshpande et al., 2001b; Gibson and Ashby, 1997). Depending on the coordination number (number of bars meeting at each node) and the balance between the number of states of self-stress and internal mechanisms, beam-based topologies can be either bending-dominated or stretching-dominated: in the former, local deformation upon global loading (e.g., compression) occurs primarily by bending of the struts (a weak deformation mode); by contrast, in the latter the vast majority of the strain energy is stored by axial deformation of the struts (Deshpande et al., 2001a). The implication is that stretching-dominated lattices are much more mechanically efficient than bending dominated lattices, with effective stiffness and yield strength scaling as $E, \sigma_y \sim (\bar{\rho})^1$ (versus $E \sim (\bar{\rho})^2$ and $\sigma_y \sim (\bar{\rho})^{1.5}$ for bending-dominated 3D lattices and stochastic foams) (Fleck, 2004). In all cases, at low relative density, strut buckling limits the strength of both stretching- and bending-dominated lattices. For nearly any strut-based lattice architecture, the effective elastic buckling strength scales with $\bar{\rho}$ as $\sigma_{el} \sim (\bar{\rho})^2$ (Ashby, 2006, 1983; Deshpande et al., 2001b), dramatically reducing mechanical performance at low relative densities. For most solid-strut topologies, the yielding-to-buckling transition occurs at a relative density of the order of ~1% (Gibson and Ashby, 1988).

Hollow-strut lattices can potentially delay the onset of buckling, shifting the yielding-to-buckling transition to $\bar{\rho}<0.1\%$ (Bauer et al., 2017; Meza et al., 2014; Pingle et al., 2011; Schaedler et al., 2011; Torrents et al., 2012; Valdevit et al., 2013, 2011). However, their mechanical properties are often much worse than analytical predictions suggest: for example $E \sim (\bar{\rho})^{1.61}$ and $\sigma_y \sim (\bar{\rho})^{1.76}$ has been experimentally extracted for hollow-strut ceramic octet lattices (Meza et al., 2014). The hollow nodes of these lattices were shown to deform through bending (regardless of the coordination number); this phenomenon, coupled with stress concentration at the nodes, causes a knock-down from classical scaling laws, which do not account for nodal topology (Bauer et al., 2016; Bauer et al., 2017; Meza et al., 2014; Portela et al., 2018; Torrents et al., 2012; Valdevit et al., 2013). Local node buckling (at low relative densities) and high sensitivity to imperfection (waviness of the strut and non-ideal strut cross section) during fabrication of these structures also contribute to the reduction of strength (Salari-Sharif et al., 2018; Valdevit et al., 2013) . Adding fillets to the nodes efficiently reduces these stress concentrations, enhancing the strength of hollow-strut lattices. Increasing node-smoothening gradually converts hollow-strut lattices to shell-based cellular materials (Han et al., 2015).

Shell-based cellular materials can be classified in closed-cell and open-cell materials, based on the connectivity of their porosity. The former include closed-cell stochastic foams (Gibson and Ashby, 1988) and 3D periodic arrangements of polygonal cells, such as closed-cell octet and cubic-octet regular foams (Berger et al., 2017) and quasi closed-cell inverse opals (do Rosário et al., 2017). While some of these materials are remarkably efficient (e.g., the nearly-isotropic cubic-octet regular foam approaches the theoretical Hashin-Shtrikman stiffness bounds (Berger et al., 2017), their topology still results in stress localization near face edges and vertices, and hence sub-optimal material utilization.

By contrast, open-cell shell-based materials can be constructed as a thin, smooth, interconnected membrane that occupies a 3D volume, with interconnected porosity. These topologies usually take the form of a locally area-minimized geometry (Meeks and Perez, 2011), and are characterized by a mean curvature approximately equal to zero everywhere on the surface. The most famous examples are triply periodic minimal surfaces (TPMS), such as the Schwarz P surface (Hyde et al., 1996; Schwarz, 1890), the Schwarz D surface (Hyde et al., 1996; Schwarz, 1890) and the gyroid surface (Schoen, 1970). The uniform curvature of these surfaces has been numerically shown to dramatically reduce local stress intensification upon macroscopic loading (Rajagopalan and Robb, 2006). Furthermore, Schwarz P shell and Schwarz D shell materials were shown to be stretching-dominated, with E and $\sigma_y$ on par with or better than most of the existing lattice architectures (Han et al., 2017, 2015; Nguyen et al., 2017). A similar class of open-cell thin shell materials was derived from simple-cubic (SC), face-centered cubic (FCC), and body-centered cubic (BCC) tube-lattices. These topologies have tailorable anisotropy and can achieve elastic moduli higher than the hollow truss lattices they are derived from (Bonatti and Mohr, 2018). Unfortunately, all these thin shell materials have two shortcomings: (i) they are significantly imperfection sensitive, as any surface roughness or waviness due to manufacturing imperfections would cause them to buckle prematurely under compression (Han et al., 2017; Hutchinson and Thompson, 2018); and (ii) they are generally difficult to fabricate with scalable manufacturing processes (inverse opal and gyroid materials can be fabricated at the nanoscale, but upscaling is challenging). The development of architected materials that can be fabricated at macroscale dimensions while still retaining dimensional control of their nano/micro-scale features is a very active research endeavor. The driving force has been the quest for scale-up of the well-known beneficial size effects on strength that metallic and ceramic materials exhibit when one or more component dimensions are reduced to the nanoscale. Excellent performance has been demonstrated in solid ceramic nanolattices (Bauer et al., 2016), hollow metallic (Schaedler et al., 2011; Zheng et al., 2016) and ceramic (Meza et al., 2014) micro and nano-lattices and nano-shell-based cellular materials (Khaderi et al., 2017, 2014). The practical applicability of all these nano-architected materials, though, is dramatically hindered by scalability challenges that derive from their manufacturing process (Bauer et al., 2017; Bishop-Moser et al., 2018). Approaches based on self-assembly, possibly combined with additive manufacturing at a larger scale, have the potential to dramatically increase scalability.

Spinodal decomposition is a near-instantaneous diffusion-driven phase transformation that converts a single-phase material into a two-phase material (one possibly being void space), with the two phases arranged in a bi-continuous topology and separated by a surface with nearly uniform negative Gaussian curvature and nearly zero mean curvature (Cahn, 1961; Jinnai et al., 1997). Spinodal decomposition may occur by a variety of physical processes, e.g., the dealloying of an Au/Ag solid solution (Hodge et al., 2007) or the heating/cooling of an emulsion gel through a critical point (Lee and Mohraz, 2010). The former results in a gold solid spinodal structure with ~50% porosity and nanoscale feature size (Fig. 1a), whereas the latter gives rise to a ~50%/50% mixture of two liquids with microscale feature size, called a bicontinuous interfacially jammed emulsion gel (bijels) (Fig. 1b). While thermodynamics favors the growth of the feature size over time, this process can be arrested (thereby controlling the length scale), by lowering the temperature (in the case of nano-porous metal) or by jamming the interface with solid nano-particles (in the case of bijels). Bijels can be used as templates for the development of structural and functional micro-architected materials, via a suite of materials conversion processes (Lee et al., 2013; Lee and Mohraz, 2011). For example, one of the liquid phases can be replaced by a photosensitive monomer, which is subsequently photopolymerized, and the other liquid phase removed, resulting in a ~50% dense polymeric cellular spinodal architecture. The surface of this cellular solid can be subsequently coated with metal or ceramic, and the polymer ultimately dissolved, with the end result being a stochastic shell-based architected material with spinodal topology (Lee and Mohraz, 2010) (Fig. 1c). This process is extremely scalable, and can result in rapid fabrication of macro-scale shell-based cellular materials with microscale feature size. The stochastic nature of this shell-based architecture, though, raises questions about its mechanical efficiency.

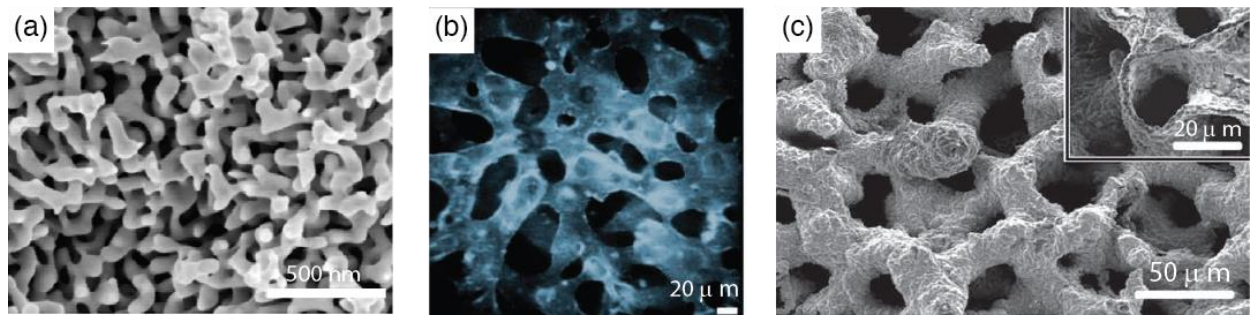

*Figure 1 – Examples of self-assembled spinodal topologies at different length scales: (a) nano-porous Au, obtained by selective etching of Au/Ag solid solutions (reproduced from (Hodge et al., 2007)); (b) spinodally decomposed bijel (reproduced from (Lee and Mohraz, 2010)); (c) micro-porous metallic shell spinodal architected material, obtained by material conversion of spinodally decomposed bijels (reproduced from (Lee and Mohraz, 2010)).*

In this work, we thoroughly investigate the mechanical response of both solid and shell-based architected materials with spinodal topologies. We numerically generate 3D spinodal cellular materials with different relative densities and characteristic feature sizes and extract shell-based spinodal cellular materials from the interface between the solid and the void space. We compute stiffness and strength under uniaxial compression by finite element simulations, which are verified by mechanical experiments conducted on polymeric samples produced by 2-photon polymerization Direct Laser Writing. We show that while solid spinodal cellular materials scale poorly with relative densities, shell spinodal architected materials are remarkably efficient, with effective Young's modulus and yield strength scaling with the relative density as $\bar{\rho}^{\,1.2\text{-}1.4}$. Importantly, spinodal shell topologies are shown to be largely imperfection insensitive, and hence robust against elastic buckling failure modes at ultra-low densities. These findings reveal that shell spinodal architected materials perform better than many available periodic topologies (both strut and shell-based), particularly at ultra-low density. These mechanical characteristics, combined with the ability to self-assemble at multiple length scales, make shell-based spinodal architected materials a remarkable class of cellular solids.

## 2. Numerical construction of spinodal topologies
### 2.1 Generation of solid spinodal topologies

Spinodal structures are formed through spinodal decomposition, a diffusion-type phase separation process that separates a compound into a mechanical mixture of phase A and phase B solid solutions by reducing the total free energy of the system (Porter et al., 1981). The phase separation processes during spinodal decomposition can numerically be described with phase field approaches (Biener et al., 2009; Crowson et al., 2009, 2007). One of the most commonly used formulations is the Cahn-Hilliard equation (Cahn and Hilliard, 1958; Sun et al., 2013), which can be written as:

$$\frac{\partial u}{\partial t} = \Delta[\frac{df(u)}{du} - \theta^2 \Delta u] \qquad (1)$$

where $u(x, y, z, t)$, is the concentration difference between the two phases (phase A and phase B) at a coordinate *(x,y,z)* at the evolution time, *t* ($u(x, y, z, t)$ is bounded between -1 and 1, with $u = -1$ denoting full phase A and $u = 1$ denoting full phase B), and $\Delta$ is the Laplacian operator. At time $t = 0$, the system is nearly homogeneous, with $u \sim 0$ everywhere. As the simulation starts, phase separation is driven by a free-energy function with a double well, which can be chosen as $f(u) = \frac{1}{4}(u^2 - 1)^2$. Regions of phase A and phase B quickly develop, with $\theta$ denoting the width of the interface between the two phases (here set to 1.1, as suggested by (Sun et al., 2013)). As time progresses the size of the single-region domains increases and the curvature of the interface decreases. For a cellular material, one of the two phases represents void space. This equation is discretized by the Finite Difference method, and solved on a cubic domain of 100 X 100 X 100 nodes, subject to periodic boundary conditions. The solution procedure is nearly identical to that reported in (Sun et al., 2013); a synopsis is reported in Appendix 1 for completeness. Solid spinodal volumes at 5 different evolution times, $t_1 - t_5$ (and hence different domain sizes and surface curvatures) and 4 different relative densities, $\bar{\rho}_1 = 20\%, \bar{\rho}_2 = 30\%, \bar{\rho}_3 = 50\%$ and $\bar{\rho}_4 = 70\%$, are extracted for post-processing (Fig. 2). Spinodal volumes were subsequently sliced in 2D images and the stack of images was imported in the commercial software Simpleware ScanIP to generate a 3D tetrahedral finite element mesh for subsequent mechanical characterization.

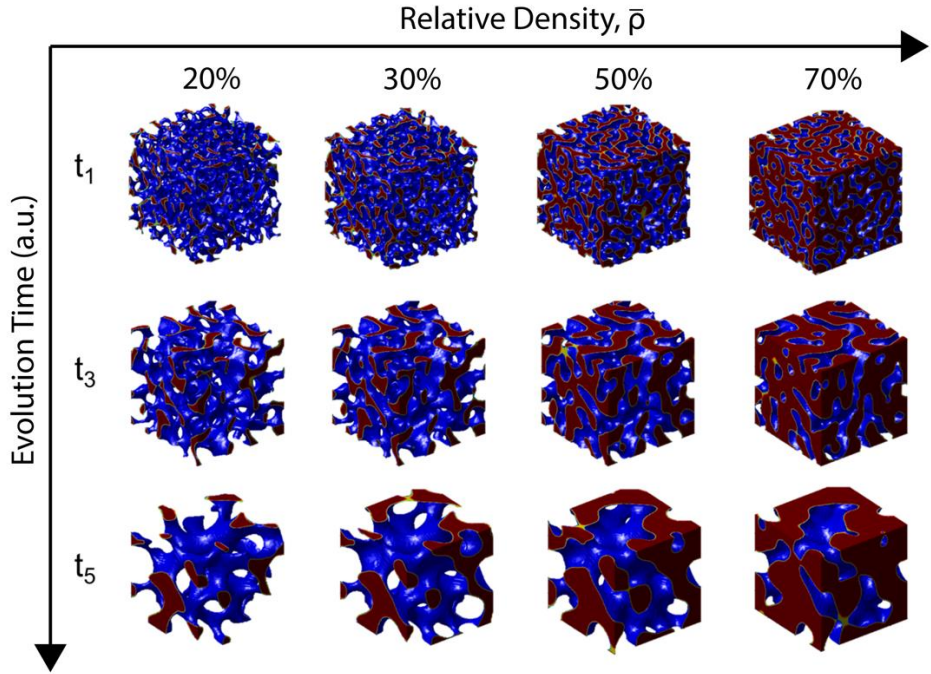

***Figure 2*** *– 3D spinodal solid models with different relative densities, $\bar{\rho} = 0.2, 0.3, 0.5,$ and $0.7$, extracted at different evolution times, $t_1$, $t_3$, and $t_5$.*

### 2.2 Calculation of the characteristic feature size

Whereas periodic architected materials have a very well-defined unit cell size, for stochastic spinodal solid this definition is fuzzier. Nonetheless, a measure of the *characteristic feature size, $\lambda$,* is necessary to enable comparison with periodic cellular materials. This dimension can be defined as the inverse of the dominant frequency in a Fast Fourier Transform (FFT) of the spinodal model (Jinnai, 2004; Kwon et al., 2010). In order to graphically illustrate the relationships between real domain and frequency domain, FFTs are initially extracted on two-dimensional slices of the solid spinodal topologies (Fig. 3). The results show three important features: (i) the intensity profiles in the FFT images are all nearly axisymmetric, indicating that all our spinodal models are isotropic; (ii) the dominant frequencies shrink as the evolution time increases, indicating that the characteristic feature size increases with evolution time; (iii) the dominant feature size is unaffected by the relative density of the solid spinodal models and only depends on evolution time, suggesting that this unique characteristic length scale can be used to uniquely describe the morphology evolution of the system (*in lieu* of time), a description that we adopt henceforth for both solid and shell spinodal topologies.

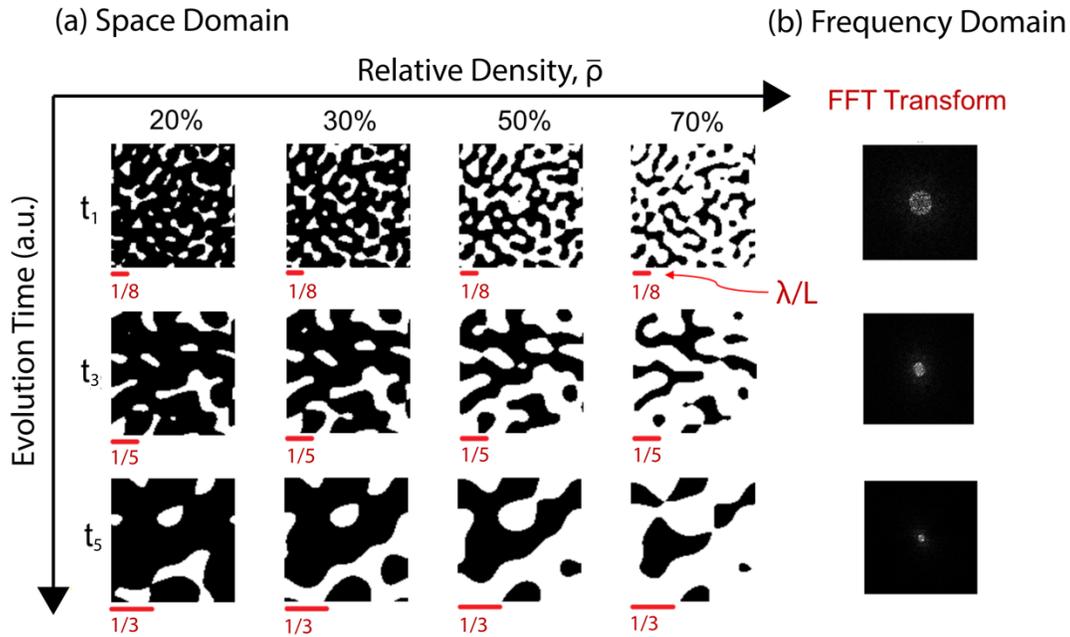

*Figure 3* – *2D slices of spinodal solid models in (a) space domain and (b) frequency domain, at different relative densities ($\bar{\rho} = 0.2, 0.3, 0.5,$ and $0.7$) and different evolution times ($t_1$, $t_3$, and $t_5$). The characteristic feature size is extracted from the FFT transforms of the models, and depends on the evolution time, but not on the relative density.*

To properly measure the characteristic feature size for each topology, FFTs are extracted for full three-dimensional spinodal models. Plots of amplitude VS frequency along the x-direction are extracted and depicted in Fig. 4a (the specific direction is immaterial, as the models were shown to be isotropic). Notice that the frequency has been multiplied by the domain size *L* in Fig. 4a; with this normalization, the x-position of the dominant frequency peak represents the number of unit cells in the domain, $N = L/\lambda$, with $\lambda$ the characteristic feature size.

Fig. 4a clearly shows that the characteristic feature size increases with evolution time, as qualitatively observed in the 2D FFTs. This time evolution is explicitly plotted in Fig. 4b, where the characteristic feature size is shown to scale with the cubic root of the evolution time, as expected by the LSW (Lifshitz-Slyozov-Wagner) theory of spinodal decomposition (Baldan, 2002; Kwon et al., 2010), for all but the first simulation. The slight disagreement on the first simulation point is attributed to an incomplete phase separation in the Cahn-Hillard evolution model (i.e., evolution is not yet fully completed at that short time, and feature growth had not fully started). This finding confirms that the numerical topologies obtained with the technique described in sec. 2.1 possess the characteristic features of naturally occurring spinodal structures.

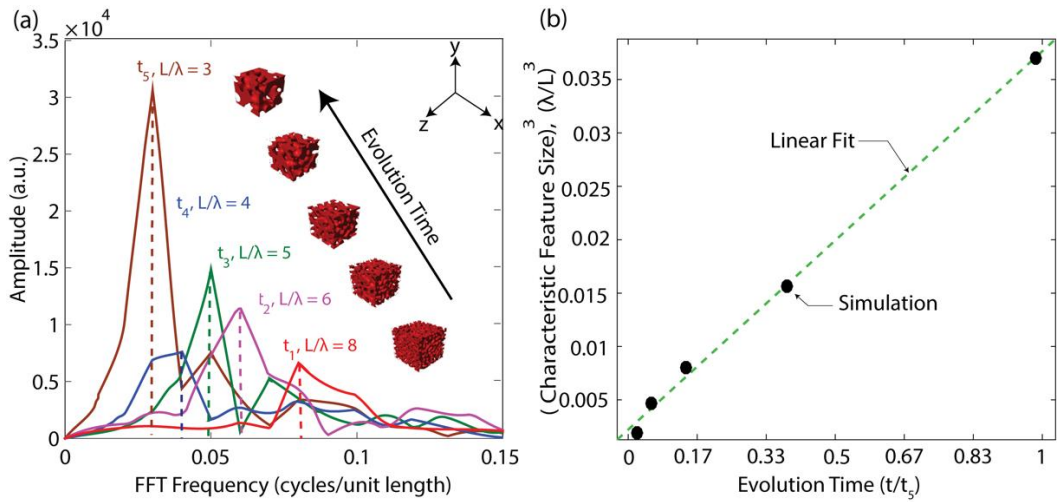

*Figure 4 – Time evolution of the inverse of the characteristic feature size, as extracted from 3D FFT analyses of solid spinodal models. (a) FFT amplitude along the x direction at five evolution times; the frequency with highest amplitude is the inverse of the characteristic feature size. (b) The characteristic feature size scales with the cubic root of the evolution time (normalized by $t_5$), as predicted by the LSW theory of spinodal decomposition.*

### 2.3 Generation of shell spinodal topologies

Spinodal shell geometries are generated by extracting the interface between solid and void phases of spinodal solid models with a three-step process: (i) the stack of 2D images for each model is imported in Matlab and a 3D matrix containing the phase information of any point in the cubic domain at a particular instant in time is generated; (ii) the Matlab 'isosurface' function is used to extract the interface between solid and void phases; (iii) the Matlab 'surface smoothening' algorithm is applied the isosurface patches, in order to remove sharp geometric transitions. The resulting three-dimensional surface is then meshed with triangular shell elements within Matlab and subsequently imported in the commercial finite element software Abaqus for further mechanical analysis. The relative density of spinodal shell topologies is tuned by controlling the shell thickness in Abaqus. With this procedure, we generate spinodal shell topologies from each of the 5 solid spinodal topologies with 50% relative density discussed above (each with a different evolution time, and hence domain size and surface curvature); three such topologies are depicted in Fig. 5a. Once the surface area per volume ratio, *S/V*, of each model is extracted, the relative density of the shell spinodal model is calculated as $\bar{\rho} = bS/V$, with *b* the shell thickness. This relationship is plotted in Fig. 5b; notice that the shell thickness has been normalized by the characteristic domain size, $\lambda$, defined in sec. 2.2 above. Using this approach, we generate shell spinodal structures with relative densities, $\bar{\rho} = 0.01\%, 0.1\%, 1\%, 5\%, 7\%$, and $10\%$ for further analysis.

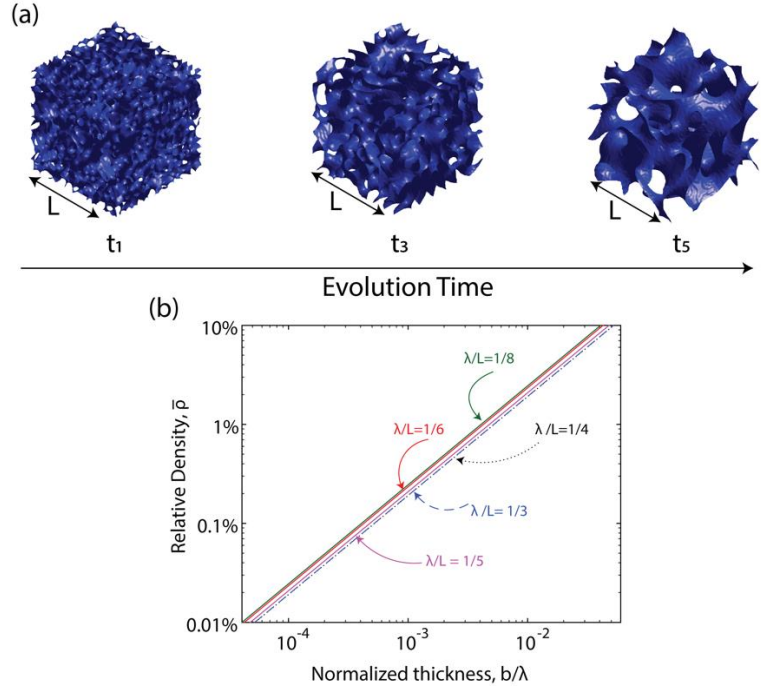

*Figure 5* – *(a) 3D spinodal shell models corresponding to the interfaces of spinodal solid models of $\bar{\rho} = 0.5$, extracted at three different evolution times ($t_1$, $t_3$, and $t_5$). (b) Relative density of shell spinodal cellular materials, as a function of the shell thickness, b, normalized by the characteristic feature size, $\lambda$. L denotes the cubic domain size.*

### 2.4 Calculation of the surface curvature

While clearly stochastic in nature, spinodal structures possess a remarkable uniform distribution of surface curvatures, with every surface patch characterized by a near-zero mean curvature and negative Gaussian curvature. Mean and Gaussian curvatures can be expressed in terms of the principal curvatures as $H = \frac{1}{2}(\kappa_1 + \kappa_2)$ and $K = \kappa_1 \cdot \kappa_2$. To confirm that the models generated in sec. 2.1 conform to this important signature of spinodal structures, we extract the principal curvatures at any location in each shell spinodal model with the algorithm described in Appendix 2. The results are plotted in Fig. 6. A number of important results clearly emerge: (i) for all topologies derived from 50% dense solid models (i.e., same amount of solid and void phases), $H\sim 0$ and $K < 0$, with most surface points within a very narrow range, as predicted by the theory of spinodal decomposition (Kwon et al., 2010); henceforth, we will use these models to construct shell spinodal topologies for mechanical characterization; (ii) for topologies derived from solid models with densities lower (higher) than 50%, $H < (>)0$ and $K < 0$, with most surface points within a very narrow range; (iii) for all topologies, the magnitudes of the principal curvatures decrease with evolution time, remaining consistent with the trends mentioned above; this is clearly consistent with the characteristic feature size coarsening over time, as noticed above.

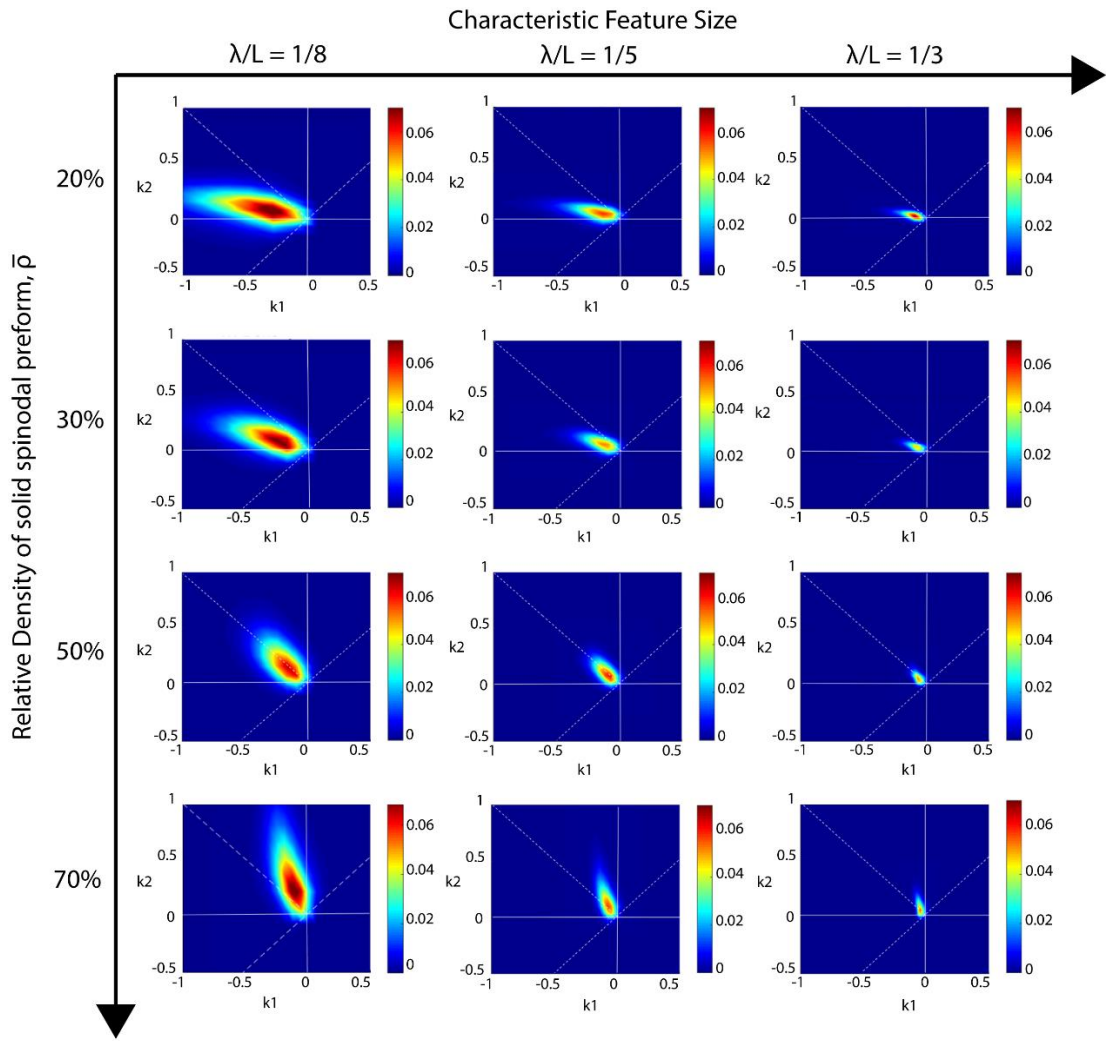

*Figure 6 – Principal curvature maps of the surface of shell spinodal models, derived from different relative densities ($\bar{\rho} = 20\%, 30\%, 50\%, and\ 70\%$) and different characteristic features sizes ($\frac{\lambda}{L} = \frac{1}{8}, \frac{1}{5}, and\ \frac{1}{3}$) of solid spinodal preforms. Notice that only the 50%-derived shell topology has near-zero mean curvature, $H = \frac{1}{2}(\kappa_1 + \kappa_2)$, throughout the surface.*

## 3. Mechanical performance of spinodal topologies

### 3.1 Numerical modeling of mechanical response

The finite element (FE) meshes of the solid and shell spinodal models, generated as described in sec. 2, are imported in the commercial FE software package Abaqus Standard, and static simulations are performed to extract the mechanical response of these cellular materials under uniaxial loading, specifically the effective Young's modulus, $E$, and the effective yield strength, $\sigma_y$. The base material is modeled as isotropic elastic-perfectly plastic, with Young's modulus, $E_s = 210\ GPa$ and yield strength, $\sigma_{ys} = 235\ MPa$ (the actual values are immaterial, as the effective properties are normalized by these values). Geometric nonlinearity is allowed during deformation.

To simulate the uniaxial response of an infinite solid, the following boundary conditions are applied on the cubic domains (Fig. 7): (i) the nodes on the –x face are restrained from moving in the x-direction but allowed to move freely in the y- and z-direction; (ii) a compressive displacement, $\delta_x = -6$, corresponding to a lattice level strain $\varepsilon_x = -6\%$, is applied to all the nodes on the +x face, which are otherwise free to move in the y- or the z-direction; (iii) all the nodes on the +y face are free to move in all directions, although their y-displacement is coupled to that of a reference node (and hence is the same for all nodes on the face); similarly, all the nodes on the +z face are free to move in all directions, although their z-displacement is coupled to that of a reference node (and hence is the same for all nodes on the face); (iv) the nodes on the –y face are constrained from moving in the y-direction but are free to move in the x- or z-direction, and the nodes on the –z face are constrained from moving in z- direction but are free to move the in x- or y- direction. Although not as rigorous as periodic boundary conditions, these boundary conditions simulate the mechanical response of infinite spinodal structures, ensuring that all plane faces remain plane and allowing for Poisson's contraction upon uniaxial compression.

The reaction force ($F_R$) integrated on all the nodes on the –x face is extracted through the simulation. The stress-strain curve for the cellular material can then be generated, with stress and strain defined as $\sigma = F_R/L^2$ and $\varepsilon = \frac{\delta_x}{L}$, respectively, with $L$ the cubic domain size. The effective Young's modulus of the cellular material, $E$, is calculated as the slope of the linear region in the stress-strain curve, and the effective yield strength, $\sigma_y$, is extracted as the 0.2% offset strength.

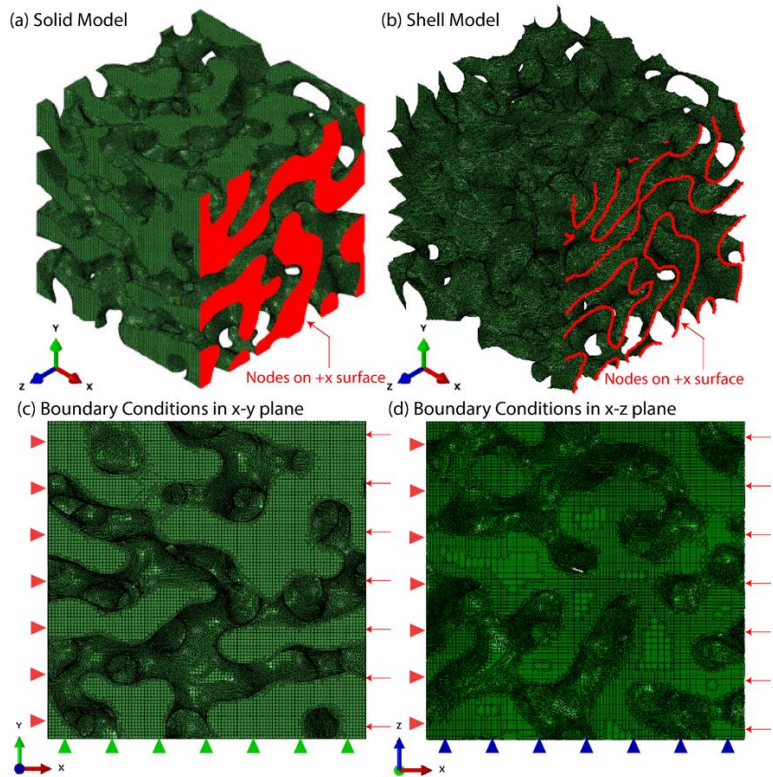

***Figure 7*** *– Finite element model of (a) a solid spinodal model ($\bar{\rho} = 50\%$) and (b) the corresponding shell spinodal model. The nodes on the positive x surface are highlighted in red indicating how the boundary conditions are applied to the models. (c-d) Boundary conditions applied to a spinodal solid model in the x-y and x-z planes. Red, green, and blue triangles indicate that translations are constrained along the x, y, and z directions, respectively. The sides opposite*

*to those indicated with red, green, and blue triangles are allowed to displace in the x, y, and z directions, respectively, but with the condition that all nodes on the side displace by the same amount. The same boundary conditions are applied to spinodal shell models.*

### 3.2 Imperfection sensitivity analysis

Thin shells (and by extension thin shell-based architected materials) are known to be stiff and strong but significantly sensitive to geometric imperfections (Hutchinson and Thompson, 2018), a characteristic that significantly limits their load carrying potential in practical applications. To quantify this effect in spinodal shell models, an imperfection sensitivity analysis is performed. First, linear eigenvalue (elastic buckling) analyses are performed on the original meshes and the first ten eigenmodes with positive eigenvalues are extracted from each mesh. Second, imperfections are introduced in the original meshes as linear combinations of the first ten eigenmodes with equal weights; the magnitude of the imperfection parameter, $\psi = h_{max}/b$, with $h_{max}$ the largest displacement from the original mesh and $b$ the shell thickness, is systematically increased in order to probe the imperfection sensitivity. Finally, post-buckling non-linear quasi-static Ricks analyses are performed in uniaxial compression, subject to same boundary conditions described above, in order to capture the deformation evolution in imperfect meshes. Figure 8 and 9 show meshes and stress-strain curves for two different 10% dense shell spinodal topologies, differing by the size of the characteristic feature, $\lambda$, relative to the domain size ((a) $\lambda/L = 1/8$, (b) $\lambda/L = 1/3$). For each topology, the response of the perfect structure ($\psi = 0$) is compared to the responses of the same topology with increasing amounts of imperfection. A very notable conclusion emerges: the topology with 8 X 8 X 8 'unit cells' in the domain (Fig. 9a) is absolutely imperfection insensitive, with imperfection magnitudes as large as 20 times the shell thickness showing no effect on the mechanical response. Even for very coarse topologies, with only 3 X 3 X 3 'unit cells' in the domain (Fig. 9b), the imperfection sensitivity becomes appreciable only when $\psi \sim 10$. (To further emphasize the imperfection insensitivity of these topologies, notice that the topology with characteristic feature size, $\frac{\lambda}{L} = 1/3$ (Fig. 8c-d and Fig. 9b) has much smaller surface area per unit volume than the topology with $\frac{\lambda}{L} = 1/8$ (Fig. 8a-b and Fig. 9a), and hence larger shell thickness; hence, in absolute terms, the imperfection magnitudes displayed in Fig. 9b are much larger than those in Fig. 9a (as evident from comparing Fig. 8a-b with Fig. 8c-d), explaining the emergence of some degree of imperfection sensitivity). Overall, the remarkable conclusion is that, unlike regular shell-based architected materials, spinodal shell based architected materials are remarkably imperfection insensitive; this important characteristic is attributed to the stochastic nature of the spinodal shell topologies. With this conclusion, all subsequent finite element analyses are performed on perfect meshed, without loss of conservatism.

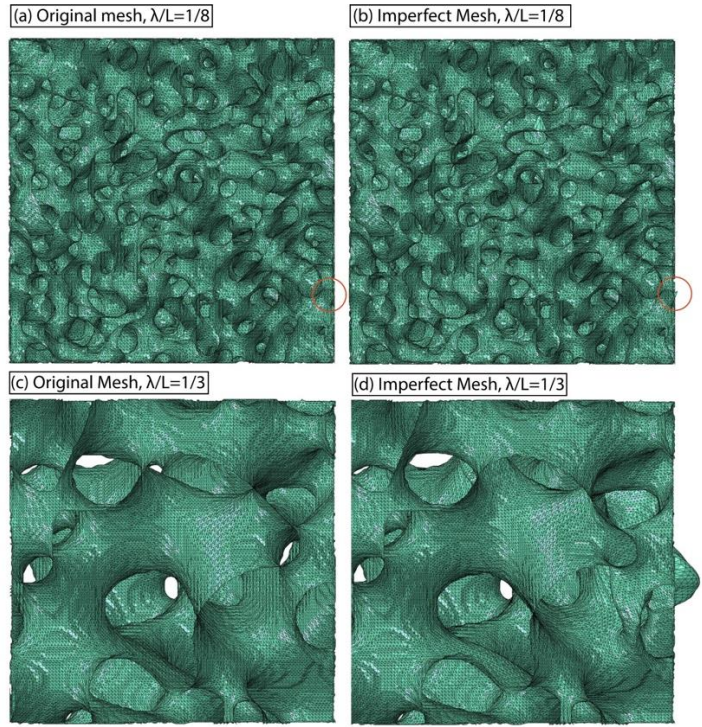

***Figure 8*** *– Comparisons of perfect (a,c) and perturbed (c,d) meshes for spinodal shell topologies with characteristic feature size, $\frac{\lambda}{L} = \frac{1}{8}$ (a,b) and $\frac{\lambda}{L} = \frac{1}{3}$ (c,d). The magnitude of the imperfection is equal to $\psi = 20$ (b) and $\psi = 9.3$ (d), respectively. Red circles in (a,b) represent examples of subtle imperfections.*

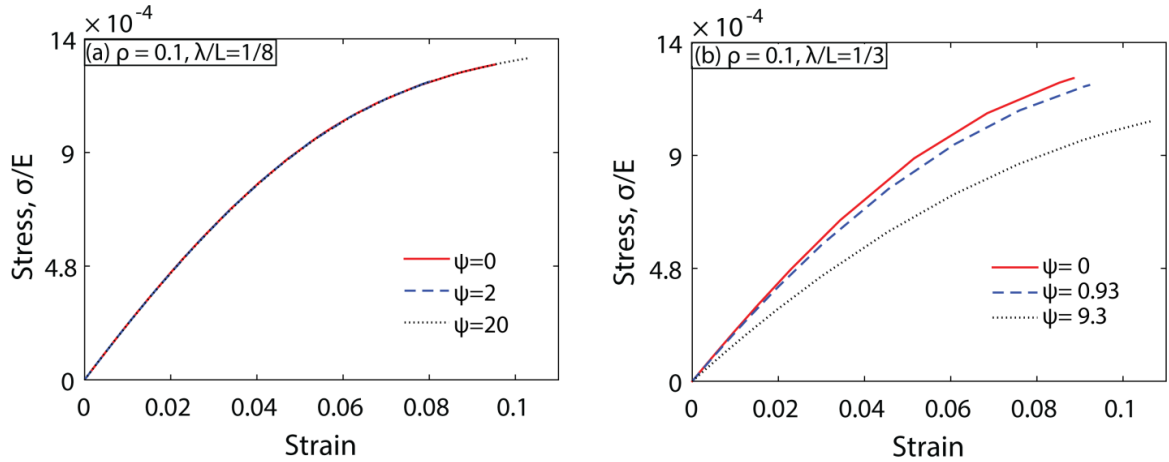

*Figure 9 – Imperfection insensitivity of shell spinodal models. The non-dimensional imperfection magnitude, ψ, is defined as the largest displacement from a perfect mesh normalized by the shell thickness. Comparison of post-buckling stress-strain curves of 10% dense spinodal shells with characteristic feature size, $\frac{\lambda}{L} = \frac{1}{8}$ (a) and for $\frac{\lambda}{L} = \frac{1}{3}$ (b), with three different magnitudes of imperfection. The constituent material has Young's modulus, $E_s = 210\ GPa$ and yield strength, $\sigma_{ys} = 235\ MPa$.*

### 3.3 Experimental verification

The accuracy of the finite element simulations is verified with a selected set of uniaxial compression experiments. Solid and shell spinodal samples were fabricated at the micro-scale via two-photon polymerization Direct Laser Writing (DLW), using a Nanoscribe Photonic Professional GT and a negative tone photoresist (IP-Dip, produced by Nanoscribe GmbH). All samples were fabricated using the Galvo mode with the 63X, N.A. 1.4 objective lens. Printing parameters were 35 mW and 17,000 µm/s for laser power and scan speed, respectively. Cubic samples with edge length, $L = 140\ \mu m$, are generated, using the solid and shell spinodal geometries calculated in Sec. 2. A characteristic feature size, $\lambda/L = 1/5$, is used for all samples. Solid spinodal samples with relative densities of 20%, 30% and 50%, and shell spinodal samples with relative densities of 5%, 7% and 10% were fabricated, respectively (all shell spinodal topologies were extracted from a 50% dense solid spinodal, as explained in Sec. 2). After DLW, samples were first immersed in propylene glycol monomethyl ether acetate (PGMEA) for 20 minutes to dissolve the remaining liquid resin, and subsequently in isopropanol for 2 minutes for final cleaning. See Fig. S1 for SEM images of all samples.

The mechanical response of all spinodal structures was characterized by displacement-controlled uniaxial compression tests with a maximum strain of 15%, performed at a constant strain rate of $0.005/s$. All tests were performed with an Alemnis Nanoindenter. Load-displacement curves were recorded and extracted, and converted to stress-strain curves as explained in Sec. 3.1. To compare experimental results with numerical predictions, finite element models of the same topologies were generated and loaded in uniaxial compression, as explained in Sec. 3.1. In order to best approximate the experimental conditions, FE simulations are performed with the bottom face fully constrained, the lateral faces free and the top face subjected to uniform vertical displacement but unconstrained laterally. In order to extract the base material properties for the numerical analyses, a number of IP-Dip cylinders with diameters between 12.5 and 50μm and aspect ratios of four were DLW printed with the same parameters as the spinodal samples and tested in uniaxial compression (see Supplementary Material Fig. S2) (ASTM D695-15, 2008). As the mechanical properties of this DLW acrylate resin are affected by the printing conditions and feature size (which cannot be kept identical for all the structures being printed), extracting an accurate stress-strain curve for the base material is challenging. To simplify the model, and in agreement with the results in Fig. S2, the base material was modeled as elastic-perfectly plastic, with Young's modulus, $E_s = 2\ GPa$, and yield strength $\sigma_{ys} = 60\ MPa$. While this simplification will slightly under-predict the ultimate strength of the structures, it is appropriate for the scope of this section.

Experimental and numerical stress-strain curves are compared in Fig. 10a-b, for solid and shell spinodal models, respectively. With the caveat on ultimate strength mentioned above, the agreement is very good throughout, thus validating all the finite element results presented in this work.

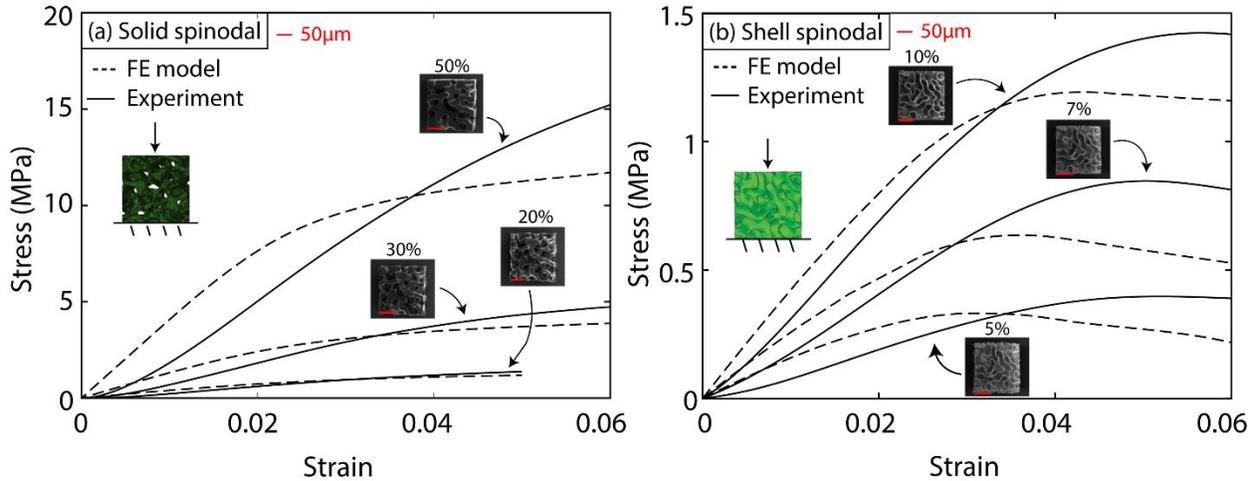

*Figure 10* – *Comparison of numerical and experimental stress-strain curves for (a) solid and (b) shell spinodal topologies at different relative densities, loaded in uniaxial compression. The characteristic feature size, $\frac{\lambda}{L} = \frac{1}{5}$ is used for all models. The insets depict SEM images of the samples, fabricated by two-photon polymerization Direct Laser Writing in IP-Dip, with a Nanoscribe GT Photonics Professional, and tested with an Alemnis Nanoindenter. Higher resolution SEM images of all samples are shown in Fig. S1. The properties of IP-Dip were measured by micropillar compression tests, and input in the simulations as an elastic-perfectly plastic solid with $E_s = 2\ GPa$, and yield strength $\sigma_{ys} = 60\ MPa$.*

### 3.4 Scaling laws and mechanical performance

Effective Young's modulus and yield strength were extracted from finite element calculations (performed as described in Sec. 3.1) for solid and shell spinodal models, for a range of relative densities, $\bar{\rho}$ and characteristic feature size, $\lambda/L$. In all simulations the constituent material was modeled as isotropic elastic-perfectly plastic, with Young's modulus, $E_s = 210\ GPa$ and yield strength, $\sigma_{ys} = 235\ MPa$. These values are representative of a relatively low yield strain metal, with yield strain $\varepsilon_y \sim 0.001$. Most results presented here are general and independent on the choice of base material.

The effective Young's modulus and yield strength of solid spinodal models depending on $\bar{\rho}$ are depicted in Fig. 11. Notice that $E \sim (\bar{\rho})^{2.0-2.6}$ and $\sigma_y \sim (\bar{\rho})^{1.7-2.3}$, with the lower exponents corresponding to topologies with larger characteristic feature size to domain size ratios. We excluded the 20% dense samples from the scaling calculations, as the 20% dense spinodal topologies displayed isolated islands on the inside of the models that depressed the mechanical properties and unrealistically steepened the slopes. While these large exponents are not unexpected at such large relative densities, this likely indicates that the solid spinodal topologies are bending dominated and the spinodal arrangement of matter is not very mechanically efficient. The numerical simulations indicate that solid spinodal models are yielding-limited (as opposed to elastic buckling-limited) for all relative densities of interest, 20-70%.

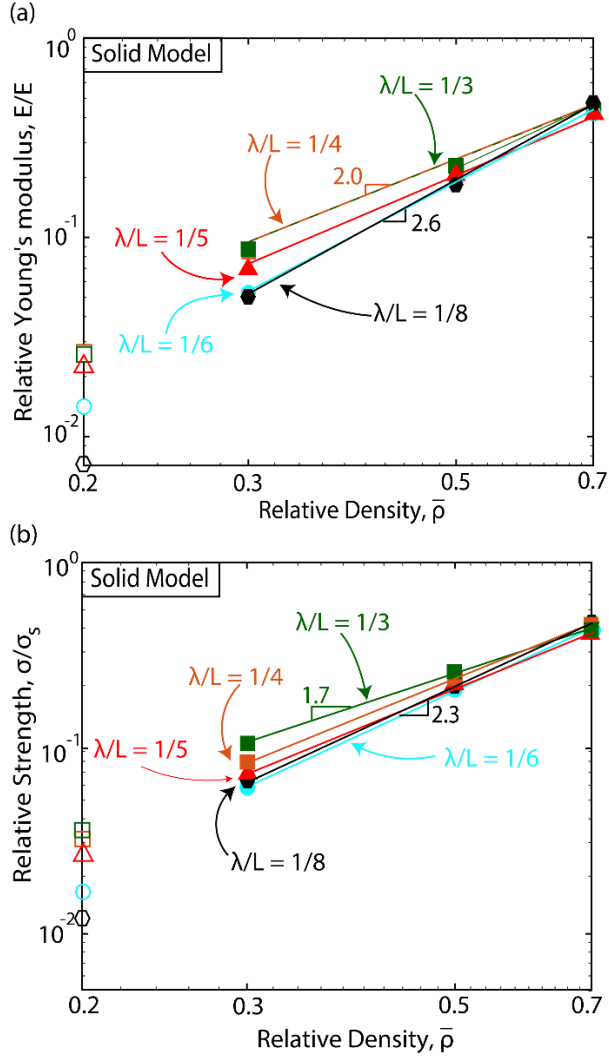

*Figure 11* – *Numerical predictions of (a) Relative Young's modulus and (b) relative yield strength of spinodal solid models, as a function of their relative density, $\bar{\rho}$. Hollow markers indicate models with isolated internal islands, which have been excluded from the scaling calculations. The constituent material has Young's modulus, $E_s = 210\ GPa$ and yield strength, $\sigma_{ys} = 235\ MPa$. Different curves refer to models with different characteristic feature size, $\lambda$. L denotes the domain size.*

Remarkably, these conclusions change dramatically for shell spinodal topologies (Fig. 12). Here, $E \sim \sigma_y \sim (\bar{\rho})^{1.2-1.3}$, a scaling that is very similar to that of the best stretching dominated lattice materials. We attribute this exceptional performance to the deformation behavior of doubly curved surfaces, which cannot be readily bent without the introduction of significant membrane stresses

– and hence deform in a stretching-dominated manner. For the base material yield strain used in the calculation, shell spinodal topologies remain yielding-limited down to relative densities in the order of 0.1%. Once the failure mechanism switches to buckling, the scaling worsens somewhat (as expected), but remains reasonably close to one. Interestingly, the relationship between scaling law exponent and characteristic feature size to domain size ratio, $\lambda/L$, is inverted compared to the case of solid spinodal topologies. In shell spinodal topologies, a larger $\lambda/L$ corresponds to a steeper scaling exponent for both stiffness and strength.

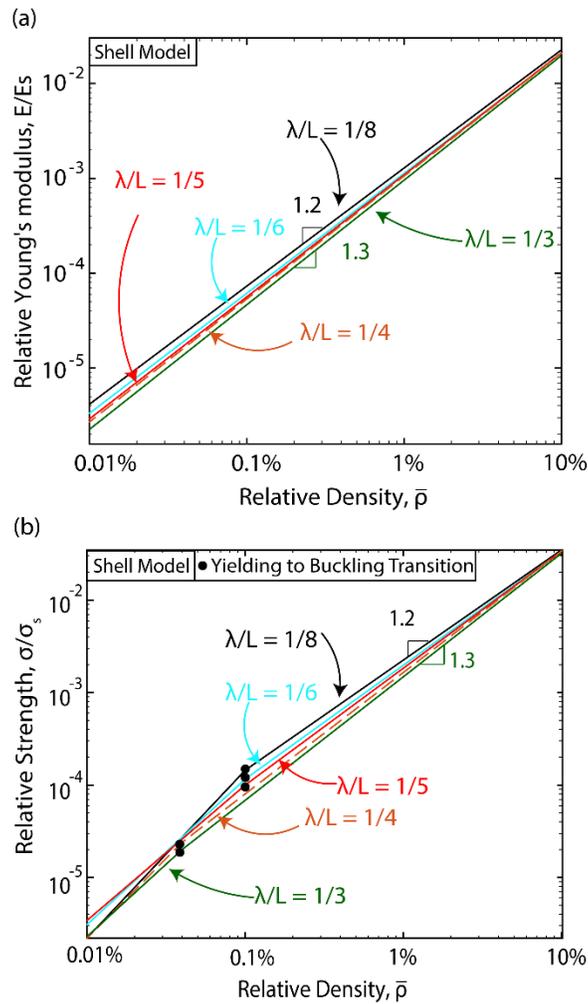

*Figure 12* – *Numerical predictions of (a) Relative Young's modulus and (b) relative yield strength of spinodal shell models, as a function of their relative density, $\bar{\rho}$. The constituent material has Young's modulus, $E_s = 210\ GPa$ and yield strength, $\sigma_{ys} = 235\ MPa$. Different curves refer to*

*models with different characteristic feature size, λ. L denotes the domain size. The black dots mark the transition from yielding to buckling failure.*

The exceptional mechanical efficiency of shell spinodal topologies is attributed to its uniform surface curvature (see Sec. 2.2), which results in a very uniform local stress distribution upon loading, avoiding areas of substantial stress intensification. To confirm this hypothesis, we contrast the local stress distribution upon uniaxial loading of a shell spinodal sample and a hollow microlattice material of the same relative densities, reproduced from (Valdevit et al., 2013) (Fig. 13); the difference in Mises stress uniformity is clearly evident. In other words, shell spinodal topologies are extremely efficient in terms of material utilization, with the entire structure yielding (or buckling) nearly simultaneously, rather than localizing failure at the nodes (a well-known shortcoming of hollow lattice materials (Meza et al., 2015, 2014; Schaedler et al., 2011; Torrents et al., 2012; Valdevit et al., 2013).

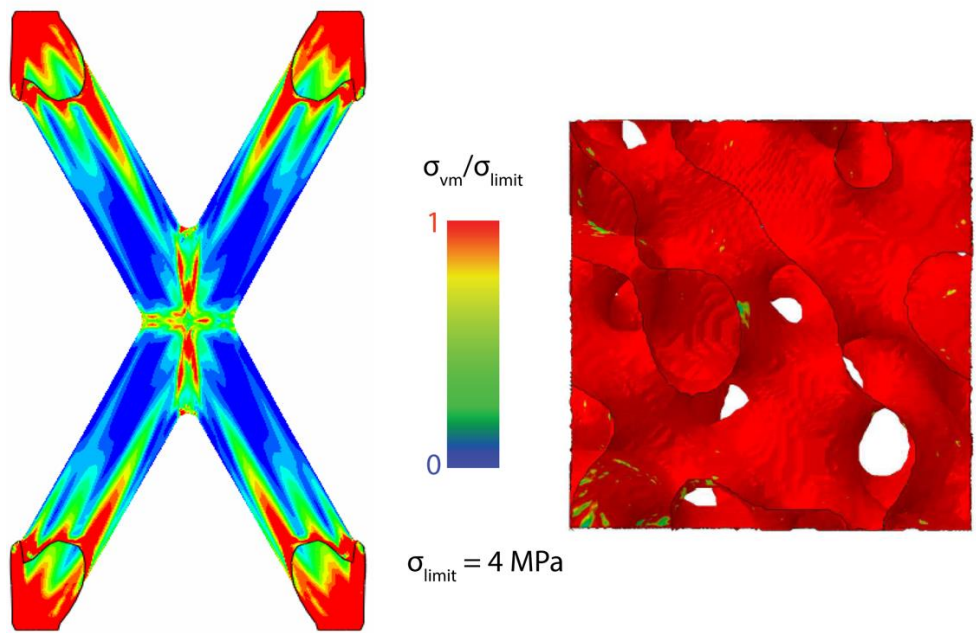

*Figure 13 – Comparison of von Mises stress distribution in a hollow microlattice unit cell with bar aspect ratio ~6, and angle $\theta = 60°$ (Valdevit et al., 2013) and a spinodal shell model with*

*characteristic feature size, $\frac{\lambda}{L} = \frac{1}{3}$, both at relative density of 0.01%, and under a compressive strain of 0.1%. The constituent material has Young's modulus, $E_s = 210\ GPa$ and yield strength, $\sigma_{ys} = 235\ MPa$. The stress limits (used to distinguish stress contours where $\frac{\sigma_{vm}}{\sigma_{limit}} \geq 1$) are colored "red", with $\sigma_{limit} = 4\ MPa$. Notice that the spinodal shell topology has a much more uniform stress distribution, yielding excellent mechanical efficiency.*

An obvious question emerges regarding the role of near-zero mean curvature on the uniformity of the stress distribution and hence the mechanical efficiency of shell spinodal topologies. As clearly illustrated in Fig. 6, all 50%-derived spinodal shell topologies possess the characteristics of minimal surfaces (i.e., $\kappa_1 = -\kappa_2$ almost everywhere on the surface), while the same is not true for shell spinodal topologies derived from solid spinodals with densities different from 50%. To quantitatively ascertain the impact of zero mean curvature on the mechanical properties of spinodal shell topologies, we generated finite element models of 20% and 70%-derived spinodal shell topologies at two different characteristic feature sizes ($\frac{\lambda}{L} = \frac{1}{8}$ and $\frac{\lambda}{L} = \frac{1}{3}$), and compared their stress-strain curve to those of 50%-derived topologies. All shell thicknesses were chosen so that all shell topologies had a 10% relative density. The results are reported in Fig. 14. Two interesting results emerge: (i) 50%-derived shell spinodal topologies outperform both the 20% and the 70%-derived topologies, at both characteristic length scales, confirming that the minimal surface characteristics of 50%-derived spinodal structures are partly responsible for their mechanical efficiency. This is attributed to a more homogeneous stress distribution (see insets). (ii) The mechanical performance of topologies derived from solid spinodal structures with densities significantly different from 50% seem to be largely unaffected by the characteristic feature size, in contrast with 50%-derived spinodal topologies.

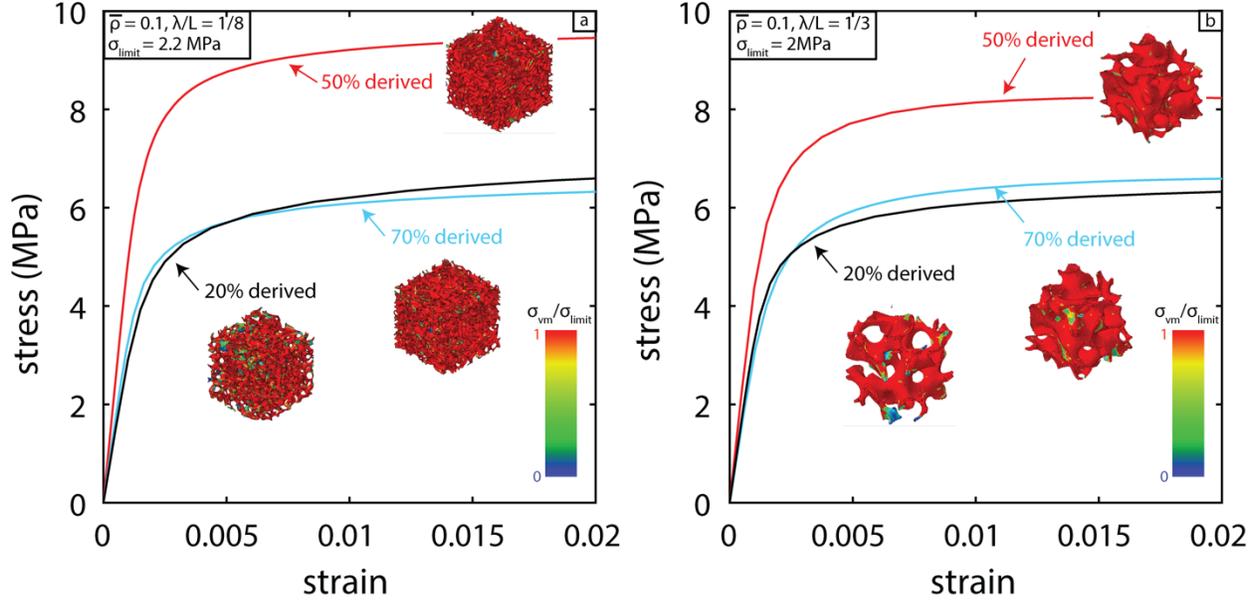

*Figure 14.* *The effect of mean surface curvature, H, on the mechanical efficiency of shell spinodal topologies. The insets display contours of von Mises stress. Notice that spinodal shell topologies derived from 50%-dense spinodal solids (and hence exhibiting near-zero mean curvature everywhere on the surface) are more efficient than spinodal shell topologies of equal relative density but derived from 20% and 70% solid spinodals (and hence having non-zero mean curvatures throughout the surface). The constituent material has Young's modulus, $E_s = 210\ GPa$ and yield strength, $\sigma_{ys} = 235\ MPa$. The stress limits (used to distinguish stress contours where $\frac{\sigma_{vm}}{\sigma_{limit}} \geq 1$) are colored "red", with (a) $\sigma_{limit} = 2.2\ MPa$ and (b) $\sigma_{limit} = 2\ MPa$.*

It is instructive to compare the mechanical efficiency of solid and shell spinodal topologies to that of a wide range of strut-based and shell-based architected materials, in particular solid strut metallic nanolattices (Khaderi et al., 2017), solid strut carbon nanolattices (Bauer et al., 2016b), hollow strut metallic microlattices (Schaedler et al., 2011; Zheng et al., 2014), hollow strut ceramic nanolattices (Meza et al., 2015, 2014), and two mechanically efficient metallic shell-based architected materials, the P-surface material (Han et al., 2015) and the D-surface material (Han et al., 2017) . These comparisons are presented in Fig. 15 in terms of relative Young's modulus, $E/E_s$, and relative yield strength, $\sigma_y/\sigma_{ys}$, versus relative density, $\bar{\rho}$. Only spinodal topologies with $\lambda/L = 1/5$ are depicted for clarity, for two different values of the base material yield strain: $\varepsilon_{ys}$=0.01 (representative of IP-Dip, thus allowing direct comparison with experimental results) and $\varepsilon_{ys}$=0.001 (representative of a metal). As all the results depicted in Fig. 15 for the 10 comparison structures are obtained experimentally, experimental results for solid and shell spinodal topologies (obtained as described in Sec. 3.2) are depicted as well. Also plotted on Fig. 15 are the Hashin-Shtrikman upper bound for stiffness (Hashin and Shtrikman, 1963; Khaderi et al., 2017) and the Suquet-Ponte-Castaneda nonlinear upper bound for strength (Castaneda and Debotton, 1992; Suquet, 1993), which represent theoretical limits for the mechanical performance of isotropic cellular materials. We emphasize that solid and shell spinodal architectures are the only isotropic topologies in Fig. 15. Notice that while spinodal solid topologies are worse than double gyroid materials and glassy carbon lattices at high relative densities, shell spinodal topologies are on par with or better than all referenced ultralight topologies across a relative density range covering 3 orders of magnitude, clearly demonstrating their exceptional mechanical efficiency.

While the model predictions for spinodal shell topologies at ultra-low density (i.e., in the buckling-limited regime) are not verified experimentally due to fabrication challenges, the remarkable imperfection insensitivity of spinodal shell topologies demonstrated in Sec. 3.2 strongly suggest that very small knock-down factors will apply, providing confidence in the numerical results.

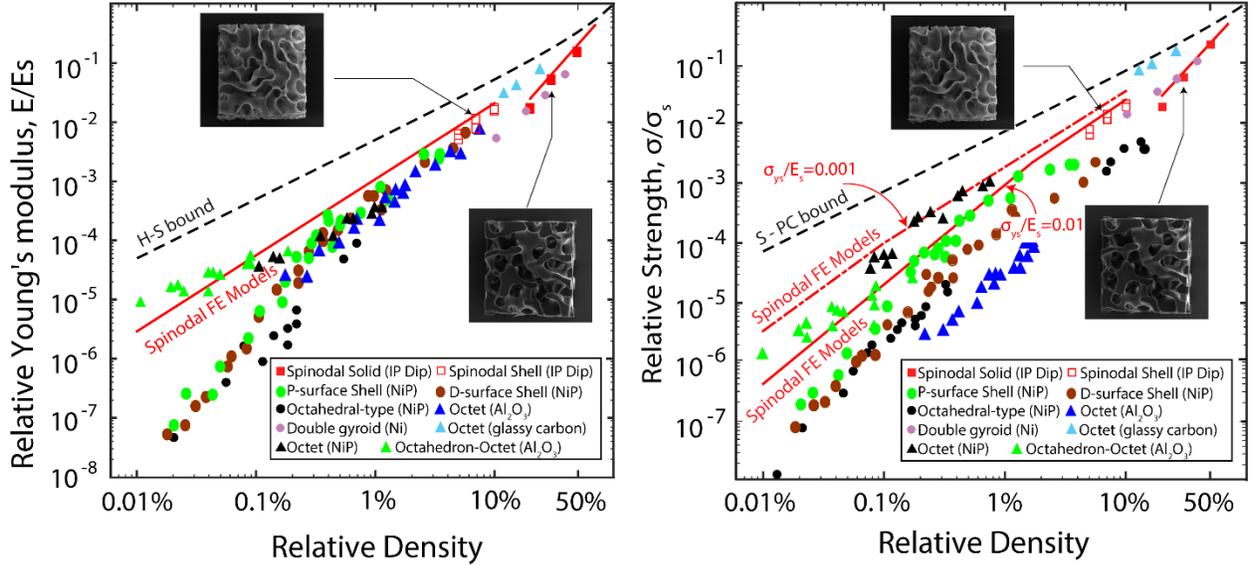

*Figure 15 – Comparison of the mechanical performance of solid and shell spinodal topologies and well-established strut-based and shell-based architected materials: hollow octahedral-type metallic microlattices (Schaedler et al., 2011), hollow octet ceramic nanolattices (Meza et al., 2014), hollow octet metallic microllatices (Zheng et al., 2014), hollow octahedron-octet nanolattices (Meza et al., 2015), solid glassy carbon nanolattices (Bauer et al., 2016b), double gyroid nanolattices (Khaderi et al., 2017), P-surface (Han et al., 2015) and D-surface (Han et al., 2017) architected materials.*

*Only spinodal topologies with characteristic feature size, $\lambda/L = 1/5$ are depicted for clarity, for two different values of the base material yield strain: $\varepsilon_{ys}$=0.01 (representative of IP-Dip, thus allowing direct comparison with experimental results) and $\varepsilon_{ys}$=0.001 (representative of a metal). Also depicted are the Hashin-Shtrikman upper bound for stiffness (Hashin and Shtrikman, 1963; Khaderi et al., 2017) and the Suquet-Ponte-Castaneda nonlinear upper bound for strength (Castaneda and Debotton, 1992; Suquet, 1993), which represent theoretical limits for the mechanical performance of isotropic cellular materials.*

## 4. Conclusions

The mechanical performance of spinodal topologies was numerically investigated and experimentally verified. Cellular materials with spinodal architectures were numerically generated by solving the Cahn-Hillard evolution equation with a Finite Difference numerical scheme. Topologies were extracted at different evolution times, corresponding to different characteristic feature size to domain size ratios, and different relative densities, i.e., different ratios of solid to void phase. We refer to these topologies as solid spinodal models. Another class of spinodal topologies was generated by retaining the interface between the solid and void phases of solid spinodal models and eliminating both volumetric phases: we refer to these topologies as shell spinodal models. The characteristic feature size of spinodal models was calculated by Fast Fourier Transform analysis, and used as a metric for 'unit cell size' of these stochastic cellular materials. This feature size was shown to scale with the cubic root of the evolution time, in agreement with the theory of spinodal decomposition. The distribution of principal curvatures of all spinodal models was also extracted, and verified to be representative of naturally occurring spinodal systems. The mechanical performance of solid and shell spinodal topologies was investigated numerically by finite element analysis. The numerical models, verified by mechanical experiments performed on solid and shell spinodal samples fabricated at the microscale by two-photon polymerization Direct Laser Writing (DLW), showed three key results: (i) shell spinodal topologies are remarkably efficient, displaying stretching dominated behavior over a very wide range of relative densities; (ii) whereas shell-based architected materials are generally imperfection sensitive, shell spinodal topologies display remarkable imperfection insensitivity, which is expected to result in superior mechanical response at ultra-low relative densities, where buckling knock-down effects plague the response of most architected materials; (iii) in the low-relative density regime ($\bar{\rho} < 1\%$) architected materials with shell spinodal topologies outperform most strut and shell-based architected materials in terms of specific stiffness and strength, while displaying complete isotropy in three dimensions.

These remarkable properties are entirely due to the topological characteristics of the architecture, and are attributed to the very uniform double curvature across the entire surface of the material, which avoids stress intensifications and prevents local bending deformations of the surface without the introduction of membrane stresses, thus imparting stretching dominated behavior to the cellular material. Importantly, base material effects have not been considered in this analysis: if emerging nano-scale fabrication techniques are used to exploit size effects in metallic and ceramic base materials (Khaderi et al., 2017; Mallory and Hajdu, 1990; Schwartzberg and Olynick, 2015), unprecedented specific strength can be achieved. These opportunities are currently under investigation.

Finally, we emphasize that the true potential of these topologies stems from the fact that they can be self-assembled, using a variety of physical processes at multiple length scales. While the experiments in this article are performed on 3D printed samples to facilitate fabrication of samples of controlled size and shape and application of mechanical loads, macroscale samples can be produced by selective etching of a metallic solid solution (Hodge et al., 2006) or by phase separation of a colloidal suspension (bijel technology) (Lee and Mohraz, 2010), resulting in nanoscale and microscale characteristic feature size, respectively. These self-assembly methodologies can be combined with additive manufacturing at a larger scale to create macroscale architected materials with unprecedented structural hierarchy, vastly improving the scalability of micro/nano-architected materials processing.

**Acknowledgments**

This work was supported by the Air Force Office of Scientific Research (AFOSR), Contract # FA9550-14-1-0352 (PM: Dr. Joycelyn Harrison). J.B. gratefully acknowledges financial support from the Deutsche Forschungsgemeinschaft (DFG), grant BA 5778/1-1. The ABAQUS Finite Element Analysis software is licensed from Dassault Systemes SIMULIA, as part of a Strategic Academic Customer Program between UC Irvine and SIMULIA. SEM imaging and *in-situ* mechanical testing were conducted at the Irvine Materials Research Institute (IMRI).The authors

**Appendix 1 – Synopsis of algorithm for spinodal decomposition**

The spinodal decomposition equation (Eq. (1)) is solved numerically with a Finite Difference scheme, as described in detail in Sun et al., 2013. Here we report the key details of the algorithm for completeness.

Equation (1) is solved on a cubic volume with edge length, *L=100*, which is discretized into a lattice of mesh size, $\ell = L/100 = 1$. Let $u_{ijk}^m$ denote the discrete value of the phase field variable $u(i,j,k,m\tau)$ at nodal point $(i,j,k)$, where $\tau$ is the integration time step ($\tau = 0.005$ is chosen as a good compromise between solution accuracy and computational cost, as suggested by (Sun et al., 2013) and *m* is the number of time steps.

A finite difference scheme is used to discretize equation (1) as:

$$\frac{u_{ijk}^{m+1} - u_{ijk}^m}{\tau} = \Delta[(u_{ijk}^m)^3 - u_{ijk}^m - \theta^2 \Delta u_{ijk}^m] \tag{A1}$$

where again $\theta$ represents the thickness of the phase A / phase B interface (here chosen as 1.1) and $\Delta$ is the Laplacian operator. Equation (2) is solved with periodic boundary conditions:

$$u(i,j,k,m\tau) = u(i+L,j,k,m\tau) \tag{A2.1}$$

$$u(i,j,k,m\tau) = u(i,j+L,k,m\tau) \tag{A2.2}$$

$$u(i,j,k,m\tau) = u(i,j,k+L,m\tau) \tag{A2.3}$$

and a randomly generated initial condition, $u(i,j,k,0) = u_0(i,j,k,0) \in [-5,5] * 10^{-4} \neq 0$.

Here, phase A ($u = -1$) represents void space, and phase B ($u = 1$) represents solid material. As the solution progresses, the system phase separates at relative early times, and subsequently continues to coarsen; during the coarsening phase, the curvature of the interface between solid and void decreases and the size of the single-phase domains increases.

To enforce a specific relative density (i.e., a prescribed ratio of phase A and phase B regions), the threshold method with the cutoff $u_c^m$ is used. The phase value of point $(i, j, k)$ at time $m\tau$ is defined as:

$$PV_{ijk}^m = H(u_{ijk}^m - u_c^m) \tag{A3}$$

where $H(u_{ijk}^m - u_c^m)$ is the Heaviside function. When $u_{ijk}^m - u_c^m > 0$, then $PV_{ijk}^m = 1$, indicating the space is occupied by material. When $u_{ijk}^m - u_c^m < 0$ then $PV_{ijk}^m = 0$ indicating void space. For a desired relative density $\bar{\rho}$, the cutoff $u_c^m$ is adjusted such that:

$$\bar{\rho} = \frac{1}{L^3}\sum_1^L\sum_1^L\sum_1^L PV_{ijk}^m = \frac{1}{L^3}\sum_1^L\sum_1^L\sum_1^L H(u_{ijk}^m - u_c^m) \tag{A4}$$

With this approach, dense spinodal topologies with arbitrary relative densities are generated.

**Appendix 2 – Algorithm for extraction of curvature probability maps**

The principal curvatures of a surface at a point $P$, $\kappa_1$ and $\kappa_2$, are defined as the minimum and maximum curvatures at that point, respectively. A paraboloid fitting method can be used to extract these curvatures at each point of the shell spinodal meshes, as proposed in (Gatzke and Grimm, 2006; Surazhsky et al., 2003). The following procedure is implemented: (i) The triangular shell mesh (generated as described in sec. 2.1) is imported in Matlab. (ii) For any node $P$ in the surface mesh, a number $J$ of elements is identified as the set of elements belonging to the three neighboring rings of node $P$, with the first neighboring ring defined as all the elements sharing node $P$, and the next two rings as the elements sharing the boundary nodes of the previous ring. (iii) Unit normal vectors to each element in the set $J$ is are constructed by normalizing the cross product of two edges of the element. (iv) The unit normal of the surface at node $P$, $\mathbf{n_P}$, is obtained as the average of the unit normals of all the elements in the first three neighboring rings of the point $P$ as follows

$$\mathbf{n_P} = \frac{1}{J}\sum_1^J \mathbf{n_e} \tag{A5}$$

where $\mathbf{n_e}$ is the unit normal of element "e". (v) All the nodes in the first three rings of $P$, along with $P$, are transformed using a rotation matrix such that $\mathbf{n_p}$ align with the z-axis of the global coordinate system. (vi) The transformed coordinates of these nodes are used to least-square fit the coefficients $a_0$ to $a_5$ of the biquadratic polynomial surface:

$$z_v = a_0 + a_1 x + a_2 y + a_3 xy + a_4 x^2 + a_5 y^2 \tag{A6}$$

(vii) The Hessian matrix of the polynomial surface is calculated as:

$$\begin{bmatrix} 2a_4 & a_3 \\ a_3 & 2a_5 \end{bmatrix} \tag{A7}$$

(viii) The pairs of minimum and maximum principal curvatures, $\kappa_1$ and $\kappa_2$, at $P$ are extracted as the eigenvalues of the Hessian matrix. (ix) The process above is repeated to find the principal curvatures at every single node of the spinodal shell mesh. Outliers in $\kappa_1$ and $\kappa_2$ are determined based on a median absolute deviation method (Leys et al., 2013), and any $(\kappa_1, \kappa_2)$ pair that has an outlier is removed. (x) Principal curvature density function maps are generated as contour plots of the probability density function $P(\kappa_1, \kappa_2)$, defined as:

$$P(\kappa_1, \kappa_2) = \frac{N(\kappa_1, \kappa_2)}{\sum N(\kappa_1, \kappa_2)} \tag{A8}$$

where $N(\kappa_1, \kappa_2)$ is the number of $(\kappa_1, \kappa_2)$ pairs having curvatures in the range $[(\kappa_1, \kappa_2), (\kappa_1 + \Delta\kappa_1, \kappa_2 + \Delta\kappa_2)]$. Here, $\Delta\kappa_1 = \frac{\max(\kappa_1) - \min(\kappa_1)}{N_1}$ and $\Delta\kappa_2 = \frac{\max(\kappa_2) - \min(\kappa_2)}{N_2}$, with $N_1$ and $N_2$ the number of intervals in $\kappa_1$ and $\kappa_2$, respectively.

This procedure was used to generate the contour plots in Fig. 6.

# References

Ashby, M.., 2006. The properties of foams and lattices. Philos. Trans. R. Soc. A Math. Phys. Eng. Sci. 364, 15–30. https://doi.org/10.1098/rsta.2005.1678

Ashby, M.F., 1983. The mechanical properties of cellular solids. Metall. Trans. A 14, 1755–1769. https://doi.org/10.1007/BF02645546

Ashby, M.F., Evans, A.G., Fleck, N.A., Gibson, L.J., Wadley, H.N.G., Hutchinson, J.W., 2000. Metal Foams : A Design Guide Metal Foams : A Design Guide. Butterworth-Heinemann, Woburn, MA, USA.

ASTM D695-15, 2008. Standard Test Method for Compressive Properties of Rigid Plastics, ASTM International. West Conshohocken, PA. https://doi.org/10.1520/D0695-15.2

Baldan, A., 2002. Progress in Ostwald ripening theories and their applications in nickel-base super alloys. J. Mater. Sci. 37, 2379–2405. https://doi.org/10.1023/A:1015408116016

Bauer, J., Meza, L.R., Schaedler, T.A., Schwaiger, R., Zheng, X., Valdevit, L., 2017. Nanolattices: An Emerging Class of Mechanical Metamaterials. Adv. Mater. 29, 1–26. https://doi.org/10.1002/adma.201701850

Bauer, J., Schroer, A., Schwaiger, R., Kraft, O., 2016a. The Impact of Size and Loading Direction on the Strength of Architected Lattice Materials. Adv. Eng. Mater. 18, 1537–1543. https://doi.org/10.1002/adem.201600235

Bauer, J., Schroer, A., Schwaiger, R., Kraft, O., 2016b. Approaching theoretical strength in glassy carbon nanolattices. Nat. Mater. 15, 438–443. https://doi.org/10.1038/nmat4561

Berger, J.B., Wadley, H.N.G., Mcmeeking, R.M., 2017. Mechanical metamaterials at the theoretical limit of isotropic elastic stiffness. Nature 1–12. https://doi.org/10.1038/nature21075

Biener, J., Wittstock, A., Zepeda-Ruiz, L.A., Biener, M.M., Zielasek, V., Kramer, D., Viswanath, R.N., Weissmüller, J., Bäumer, M., Hamza, A. V., 2009. Surface-chemistry-driven actuation in nanoporous gold. Nat. Mater. 8, 47–51. https://doi.org/10.1038/nmat2335

Bishop-Moser, J., Spadaccini, C.M., Andres, C., 2018. Metamaterials Manufacturing - Pathways to Industrial Competitiveness. Ann Arbor, MI.

Bonatti, C., Mohr, D., 2018. Smooth-shell metamaterials of cubic symmetry: Anisotropic elasticity, yield strength and specific energy absorption. Acta Mater. 164, 301–321. https://doi.org/10.1016/j.actamat.2018.10.034

Cahn, J.W., 1961. On spinodal decomposition. Acta Metall. 9, 795–801. https://doi.org/10.1016/0001-6160(61)90182-1

Cahn, J.W., Hilliard, J.E., 1958. Free Energy of a Nonuniform System. I. Interfacial Free Energy. J. Chem. Phys. 28, 258–267. https://doi.org/10.1063/1.1744102

Castaneda, P.P., Debotton, G., 1992. On the Homogenized Zield Strength of Two-Phase Composites. Proc. R. Soc. Lond. A 438, 419–431. https://doi.org/10.1021/bi051624q

Crowson, D.A., Farkas, D., Corcoran, S.G., 2009. Mechanical stability of nanoporous metals with small ligament sizes. Scr. Mater. 61, 497–499. https://doi.org/10.1016/j.scriptamat.2009.05.005

Crowson, D.A., Farkas, D., Corcoran, S.G., 2007. Geometric relaxation of nanoporous metals: The role of surface relaxation. Scr. Mater. 56, 919–922. https://doi.org/10.1016/j.scriptamat.2007.02.017

Deshpande, V.S., Ashby, M.F., Fleck, N.A., 2001a. Foam topology: Bending versus stretching dominated architectures. Acta Mater. 49, 1035–1040. https://doi.org/10.1016/S1359-6454(00)00379-7

Deshpande, V.S., Fleck, N.A., Ashby, M.F., 2001b. Effective properties of the octet-truss lattice material. J. Mech. Phys. Solids 49, 1747–1769. https://doi.org/10.1016/S0022-5096(01)00010-2

do Rosário, J.J., Berger, J.B., Lilleodden, E.T., McMeeking, R.M., Schneider, G.A., 2017. The stiffness and strength of metamaterials based on the inverse opal architecture. Extrem. Mech. Lett. 12, 86–96. https://doi.org/10.1016/j.eml.2016.07.006

Fleck, N. a, 2004. An overview of the mechanical properties of foams and periodic lattice materials. Cell. Met. Polym. 2004 1–4.

Fleck, N.A., Deshpande, V.S., Ashby, M.F., 2010. Micro-architectured materials: Past, present and future. Proc. R. Soc. A Math. Phys. Eng. Sci. 466, 2495–2516. https://doi.org/10.1098/rspa.2010.0215

Gatzke, T.D., Grimm, C.M., 2006. Estimating Curvature on Triangular Meshes. Int. J. Shape Model. 12, 1–28. https://doi.org/10.1142/S0218654306000810

Gibson, L.J., Ashby, M.F., 1997. Cellular solids, 2nd ed. Cambridge University Press, Cambridge, MA.

Gibson, L.J., Ashby, M.F., 1988. Cellular Solids: Structure and Properties. Pergamon Press, Oxford, United Kingdom.

Han, S.C., Choi, J.M., Liu, G., Kang, K., 2017. A microscopic shell structure with schwarz's D-Surface. Sci. Rep. 7, 1–8. https://doi.org/10.1038/s41598-017-13618-3

Han, S.C., Lee, J.W., Kang, K., 2015. A New Type of Low Density Material: Shellular. Adv. Mater. 27, 5506–5511. https://doi.org/10.1002/adma.201501546

Hashin, Z., Shtrikman, S., 1963. A variational approach to the theory of the elastic behaviour of multiphase materials. J. Mech. Phys. Solids 11, 127–140. https://doi.org/10.1016/0022-

5096(63)90060-7

Hodge, A.M., Biener, J., Hayes, J.R., Bythrow, P.M., Volkert, C.A., Hamza, A. V., 2006. Scaling equation for yield strength of nanoporous open-cell foams. Acta Mater. 55, 1343–1349. https://doi.org/10.1016/j.actamat.2006.09.038

Hutchinson, J.W., Thompson, J.M.T., 2018. Imperfections and Energy Barriers in Shell Buckling. Int. J. Solids Struct. 1–32.

Hyde, S., Landh, T., Lidin, S., Ninham, B.W., Andersson, S., Larsson, K., 1996. The Language of Shape. Elsevier, Danvers, Ma.

Jinnai, H., Koga, T., Nishikawa, Y., Hashimoto, T., Hyde, S., 1997. Curvature Determination of Spinodal Interface in a Condensed Matter System. Phys. Rev. Lett. 78, 2248–2251. https://doi.org/10.1103/PhysRevLett.78.2248

Jinnai, H.Y.Ni.T.I.T.Ni., 2004. Emerging Technologies for the 3D Analysis of Polymer Structures. Adv. Polym. Sci. 171, 137–194. https://doi.org/10.1007/b13016

Khaderi, S.N., Deshpande, V.S., Fleck, N.A., 2014. The stiffness and strength of the gyroid lattice. Int. J. Solids Struct. 51, 3866–3877. https://doi.org/10.1016/j.ijsolstr.2014.06.024

Khaderi, S.N., Scherer, M.R.J., Hall, C.E., Steiner, U., Ramamurty, U., Fleck, N.A., Deshpande, V.S., 2017. The indentation response of Nickel nano double gyroid lattices. Extrem. Mech. Lett. 10, 15–23. https://doi.org/10.1016/j.eml.2016.08.006

Kwon, Y., Thornton, K., Voorhees, P.W., 2010. Morphology and topology in coarsening of domains via non-conserved and conserved dynamics. Philos. Mag. 90, 317–335. https://doi.org/10.1080/14786430903260701

Lee, M.N., Mohraz, A., 2011. Hierarchically porous silver monoliths from colloidal bicontinuous interfacially jammed emulsion gels. J. Am. Chem. Soc. 133, 6945–6947.

https://doi.org/10.1021/ja201650z

Lee, M.N., Mohraz, A., 2010. Bicontinuous macroporous materials from bijel templates. Adv. Mater. 22, 4836–4841. https://doi.org/10.1002/adma.201001696

Lee, M.N., Thijssen, J.H.J., Witt, J.A., Clegg, P.S., Mohraz, A., 2013. Making a robust interfacial scaffold: Bijel rheology and its link to processability. Adv. Funct. Mater. 23, 417–423. https://doi.org/10.1002/adfm.201201090

Leys, C., Ley, C., Klein, O., Bernard, P., Licata, L., 2013. Detecting outliers: Do not use standard deviation around the mean, use absolute deviation around the median. J. Exp. Soc. Psychol. 49, 764–766. https://doi.org/10.1016/j.jesp.2013.03.013

Mallory, G.O., Hajdu, J.B., 1990. Electroless Plating: Fundamentals and Applications, 1st ed. William Andrew Publishing, New York, USA.

Meeks, W.H., Perez, J., 2011. The classical theory of minimal surfaces. Bull. Amer. Math. Soc. 48, 325–407. https://doi.org/10.1090/S0273-0979-2011-01334-9

Meza, L.R., Das, S., Greer, J.R., 2014. Strong, lightweight, and recoverable three-dimensional. Science (80-. ). 345, 1322–1326. https://doi.org/10.1126/science.1255908

Meza, L.R., Zelhofer, A.J., Clarke, N., Mateos, A.J., Kochmann, D.M., Greer, J.R., 2015. Resilient 3D hierarchical architected metamaterials. Proc. Natl. Acad. Sci. 112, 11502–11507. https://doi.org/10.1073/pnas.1509120112

Nguyen, B.D., Han, S.C., Jung, Y.C., Kang, K., 2017. Design of the P-surfaced shellular, an ultra-low density material with micro-architecture. Comput. Mater. Sci. 139, 162–178. https://doi.org/10.1016/j.commatsci.2017.07.025

Pingle, S.M., Fleck, N.A., Deshpande, V.S., Wadley, H.N.G., 2011. Collapse mechanism maps for the hollow pyramidal core of a sandwich panel under transverse shear. Int. J. Solids Struct.

48, 3417–3430. https://doi.org/10.1016/j.ijsolstr.2011.08.004

Portela, C.M., Greer, J.R., Kochmann, D.M., 2018. Impact of node geometry on the effective stiffness of non-slender three-dimensional truss lattice architectures. Extrem. Mech. Lett. 22, 138–148. https://doi.org/10.1016/j.eml.2018.06.004

Porter, D.A., Easterling, K.E., Sherif, M.Y., 1981. Phase Transformations in Metals and Alloys. CRC Press.

Rajagopalan, S., Robb, R.A., 2006. Schwarz meets Schwann: Design and fabrication of biomorphic and durataxic tissue engineering scaffolds. Med. Image Anal. 10, 693–712. https://doi.org/10.1016/j.media.2006.06.001

Salari-Sharif, L., Godfrey, S.W., Tootkaboni, M., Valdevit, L., 2018. The effect of manufacturing defects on compressive strength of ultralight hollow microlattices: A data-driven study. Addit. Manuf. 19, 51–61. https://doi.org/10.1016/j.addma.2017.11.003

Schaedler, T.A., Jacobsen, A.J., Torrents, A., Sorensen, A.E., Lian, J., Greer, J.R., Valdevit, L., Carter, W.B., 2011. Ultralight Metallic Microlattices. Science (80-. ). 334, 962–965. https://doi.org/10.1126/science.1211649

Schoen, A.H., 1970. Infinite Periodic Minimal Surfaces without Self-Intersections, NASA Technical Note TN D-5541.

Schwartzberg, A.M., Olynick, D., 2015. Complex Materials by Atomic Layer Deposition. Adv. Mater. 27, 5778–5784. https://doi.org/10.1002/adma.201500699

Schwarz, H.A., 1890. Gesammelte Mathematische Abhandlungen. Springer, Berlin.

Sun, X.-Y., Xu, G.-K., Li, X., Feng, X.-Q., Gao, H., 2013. Mechanical properties and scaling laws of nanoporous gold. J. Appl. Phys. 113, 023505. https://doi.org/10.1063/1.4774246

Suquet, P.M., 1993. Overall potentials and extremal surfaces of power law or ideally plastic

composites. J. Mech. Phys. Solids 41, 981–1002. https://doi.org/10.1016/0022-5096(93)90051-G

Surazhsky, T., Magid, E., Soldea, O., Elber, G., Rivlin, E., 2003. A Comparison of Gaussian and Mean Curvatures Estimation Methods on Triangular Meshes. 2003 IEEE Int. Conf. Robot. Autom. (Cat. No.03CH37422) 1, 1021–1026. https://doi.org/10.1109/ROBOT.2003.1241726

Torrents, A., Schaedler, T.A., Jacobsen, A.J., Carter, W.B., Valdevit, L., 2012. Characterization of nickel-based microlattice materials with structural hierarchy from the nanometer to the millimeter scale. Acta Mater. 60, 3511–3523. https://doi.org/10.1016/j.actamat.2012.03.007

Valdevit, L., Godfrey, S.W., Schaedler, T.A., Jacobsen, A.J., Carter, W.B., 2013. Compressive strength of hollow microlattices: Experimental characterization, modeling, and optimal design. J. Mater. Res. 28, 2461–2473. https://doi.org/10.1557/jmr.2013.160

Valdevit, L., Jacobsen, A.J., Greer, J.R., Carter, W.B., 2011. Protocols for the optimal design of multi-functional cellular structures: From hypersonics to micro-architected materials. J. Am. Ceram. Soc. 94, 15–34. https://doi.org/10.1111/j.1551-2916.2011.04599.x

Zheng, X., Smith, W., Jackson, J., Moran, B., Cui, H., Chen, D., Ye, J., Fang, N., Rodriguez, N., Weisgraber, T., Spadaccini, C.M., 2016. Multiscale metallic metamaterials. Nat. Mater. 15, 1100–1106. https://doi.org/10.1038/nmat4694

Zheng, X.Y., Lee, H., Weisgraber, T.H., Shusteff, M., Deotte, J., Duoss, E.B., Kuntz, J.D., Biener, M.M., Ge, Q., Jackson, J.A., Kucheyev, S.O., Fang, N.X., Spadaccini, C.M., 2014. Ultralight, ultrastiff mechanical metamaterials. Science (80-. ). 344, 1373–1378. https://doi.org/10.1126/science.1252291

**Supplementary Material**

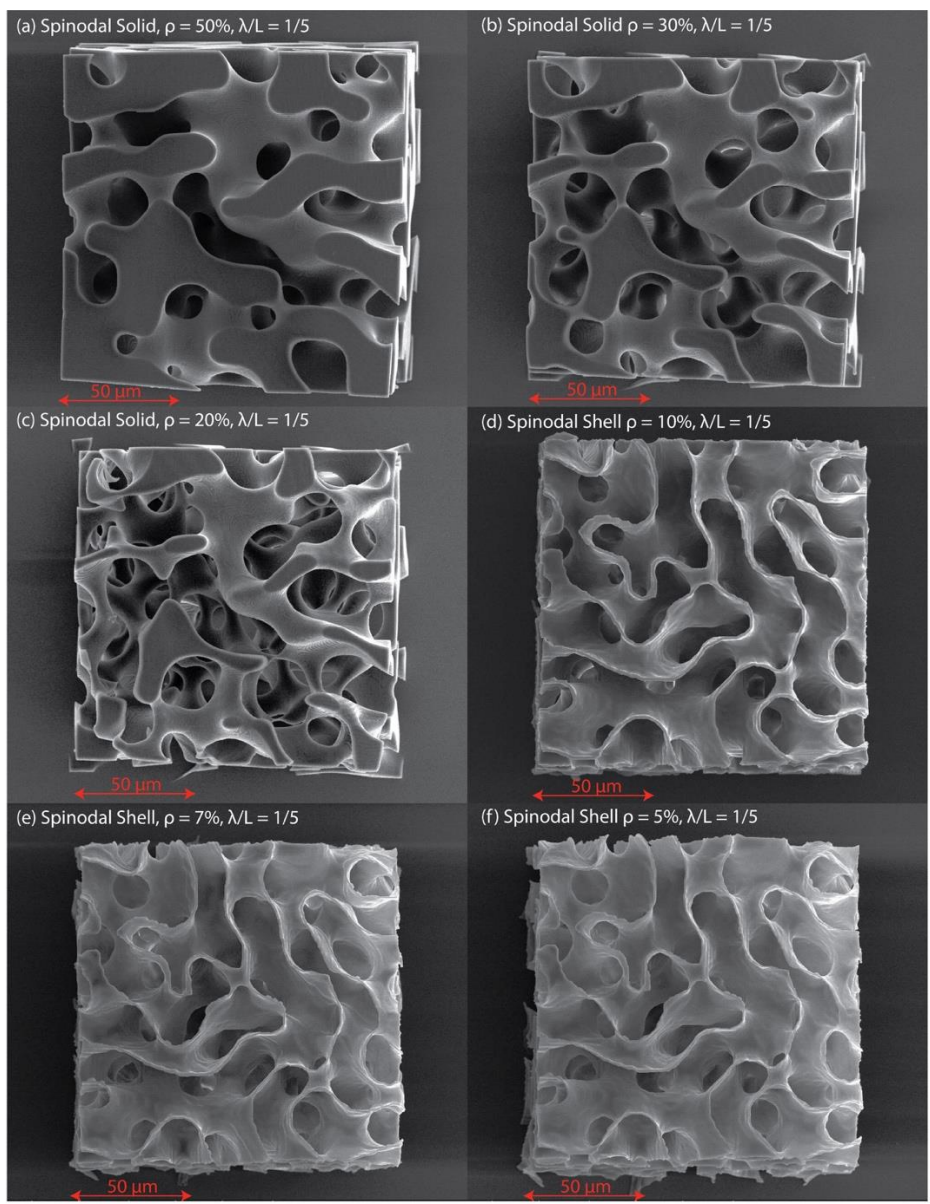

*Fig. S1. SEM images of spinodal samples fabricated in IP-Dip by two-photon polymerization Direct Laser Writing. All samples have characteristic feature size, $\frac{\lambda}{L} = \frac{1}{5}$, with L the domain size. (a) Spinodal solid model, $\overline{\rho} = 50\%$. (b) Spinodal solid model, $\overline{\rho} = 30\%$. (c) Spinodal solid model, $\overline{\rho} = 20\%$. (d) Spinodal shell model, $\overline{\rho} = 10\%$. (e) Spinodal shell model, $\overline{\rho} = 7\%$. (f) Spinodal shell model, $\overline{\rho} = 5\%$.*

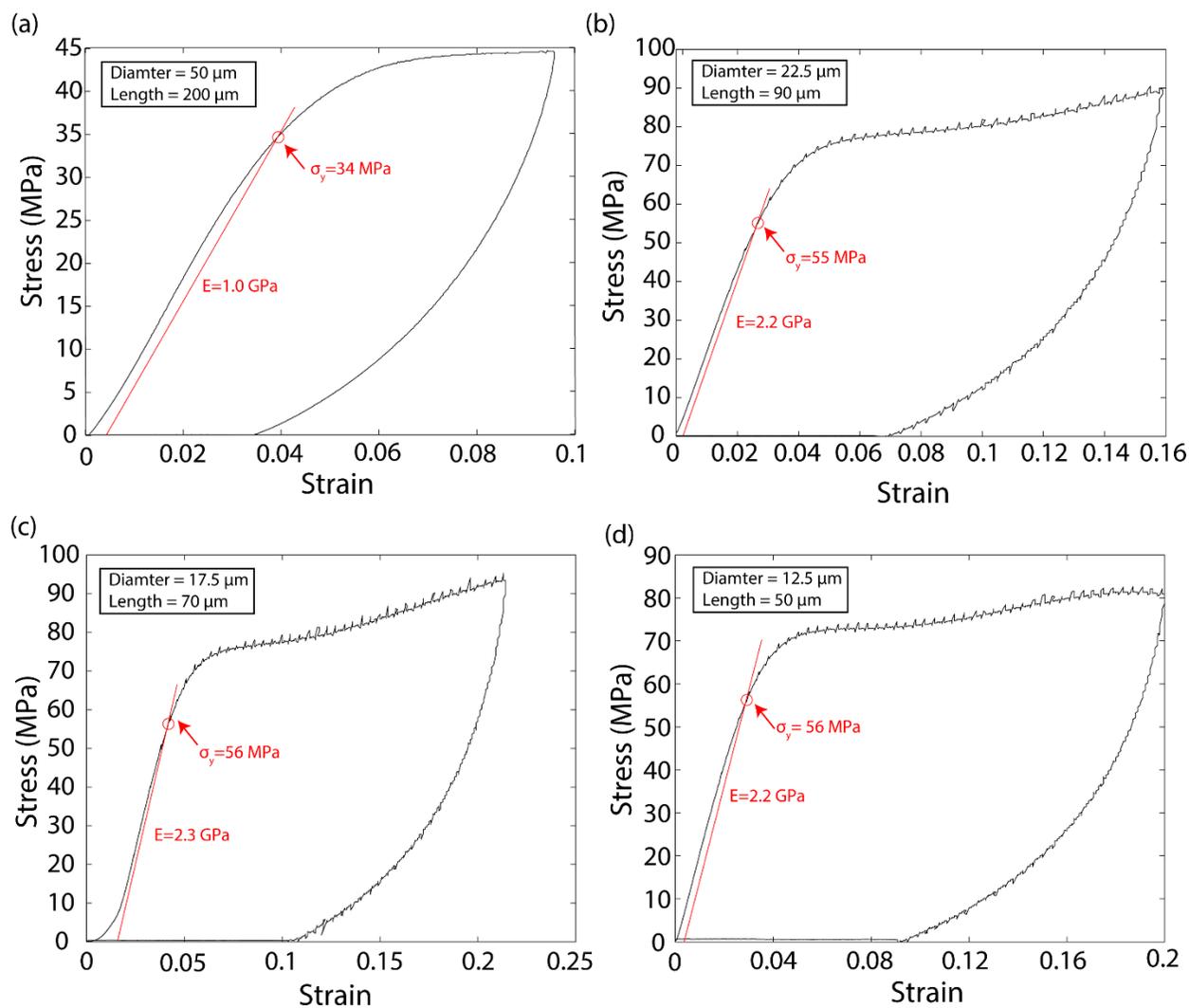

*Fig. S2.* *Uniaxial response of the material IP-Dip, measured by compression experiments on micro-pillars of aspect ratio of 4 and different diameters, manufactured by two-photon polymerization Direct Laser Writing with a Nanoscribe Photonics GT Professional.*